\documentstyle[preprint,aps]{revtex}

\def\ni{\noindent}

\def\e{\varepsilon}
\def\d{\delta}

\def\ee{\vspace{3mm}}
\def\nl{\vspace*{-2.5mm}\newline}
\def\ii{\hspace*{10mm}}
\def\ni{\noindent}
\def\-{\leftarrow}
\def\xor{\oplus}
\def\maj{\mbox{MAJ}}
\def\Hs{H_{\hspace*{-0.4mm}\mbox{\tiny\sl S}}}
\def\Hvn{H_{\hspace*{-0.5mm}\mbox{\tiny\sl VN}}}
\def\S{{\cal S}}
\def\carrow{\mbox{$\leftarrow\!\!\!\!\subset\,$}}
\def\crarrow{\mbox{$\leftarrow\!\!\!\!\subset_R\,$}}
\def\ciarrow{\mbox{$\leftarrow\!\!\!\!<\,$}}

\begin{document}

\draft

\preprint{schums1.tex; submitted to Phys. Rev. A 3/7/96}

\title{Schumacher's quantum data compression \\ as a quantum computation}

\author{Richard Cleve}
\address{Department of Computer Science,\\
University of Calgary,\\ Calgary, Alberta, Canada T2N 1N4.}
\author{David P. DiVincenzo}
\address{
IBM Research Division\\
Thomas J. Watson Research Center\\
P. O. Box 218\\
Yorktown Heights, NY 10598 USA\\
}

\date{\today}

\maketitle

\begin{abstract}
An explicit algorithm for performing Schumacher's noiseless
compression of quantum bits is given.  This algorithm is based on a
combinatorial expression for a particular bijection among binary
strings.  The algorithm, which adheres to the rules of reversible
programming, is expressed in a high-level pseudocode language.  It is
implemented using $O(n^3)$ two- and three-bit primitive reversible
operations, where $n$ is the length of the qubit strings to be
compressed.  Also, the algorithm makes use of $O(n)$ auxiliary qubits;
however, space-saving techniques based on those proposed by Bennett
are developed which reduce this workspace to $O(\sqrt{n})$ while
increasing the running time by less than a factor of two.
\end{abstract}

\pagebreak

\section{Introduction}
\label{sec:one}

There is considerable interest in the controlled generation,
manipulation and transportation of individual quantum states;
applications of such resources are envisioned in new kinds of data
transmission, cryptography and computation.  The quantum extension of
conventional bits, called {\em qubits}, have been subject to
considerable exploration lately.  A single qubit is embodied in the
state of a single two-state quantum system, such as the spin degree of
freedom of an electron or other spin-$1 \over 2$ particle, where the
spin-up state of the particle is denoted by $|0\rangle$ and the
spin-down state is denoted by $|1\rangle$.  The basic laws of quantum
physics dictate that a description of the entire possible state-space
of the qubit is given by the wavefunction
\begin{equation} |\Psi\rangle = \alpha |0\rangle+\beta|1\rangle,
\end{equation}
where $\alpha$ and $\beta$ are any two
complex numbers such that $|\alpha |^2+|\beta |^2=1$. This is called a
``qubit'' since it can assume one of two binary values, but of course
it has fundamentally different properties because of the possibility
of it being in a superposition of these two values.  The properties with
which quantum mechanics endows the qubit make possible a kind of
cryptography which is fundamentally secure against eavesdropping
attacks\cite{SciAm}, and computations which apparently violate the
complexity-class categorizations for ordinary boolean computers\cite{EJ}.

One of the ideas of this sort that has been understood recently is the
possibility of data compression for qubits.  In classical information
theory, if $n$ bits, $x_1,\ldots,x_n$, are each sampled independently
according to some probability distribution $p = (p_0,p_1)$ (on the set
$\{0,1\}$) then the string $x_1\ldots x_n$ may be compressed to a $n
\Hs(p)$-bit string (where $\Hs(p) = - \sum_{i=0}^1 p_i \log p_i$, the
Shannon entropy \cite{Shan})---and no further---in the following
asymptotic sense.  For any $\e, \d > 0$, for sufficiently large $n$,
for any $\lambda(n) \ge n (\Hs(p) + \d)$, $\lambda(n) \in
\{1,\ldots,n\}$, there exists a compression scheme that compresses
$x_1\ldots x_n$ to $y_1\ldots y_{\lambda(n)}$, and such that
$x_1\ldots x_n$ can be successfully recovered from $y_1\ldots
y_{\lambda(n)}$ with probability greater than $1 - \e$.  Moreover, the
above compression is the maximum possible in the sense that, for any
$\e, \d > 0$, for sufficiently large $n$, for any $\lambda(n) \le n
(\Hs(p) - \d)$, for any compression scheme that maps $x_1\ldots x_n$
to $y_1\ldots y_{\lambda(n)}$, the probability that $x_1\ldots x_n$
can be successfully recovered from $y_1\ldots y_{\lambda(n)}$ is less
than $\e$.

The quantum physical analogue of the above scenario involves the compression
of a string of qubits, instead of bits.
Note that there are a continuum of possible states for  each qubit, rather
than two possible values.
We shall consider the ``discrete'' case, where a probability distribution
is concentrated on some finite set of qubit states
$\S = \{|\Psi_1\rangle,\ldots,|\Psi_m\rangle\}$.
Let the respective probabilities be $p = (p_1,\ldots,p_m)$.
In the language
of quantum physics, $(\S,p)$ defines an {\em ensemble} of states.  Let
$|\alpha_1\rangle \ldots |\alpha_n\rangle$ be a string of $n$ qubits,
each sampled independently from $(\S,p)$.
Define a {\it compressor A} as a unitary transformation that maps
$n$-qubit strings to $n$-qubit strings.  Again let $\lambda(n) \in
\{1,\ldots,n\}$.  It is to be understood that, on input
$|\alpha_1\rangle \ldots |\alpha_n\rangle$, the first $\lambda(n)$
qubits that are output by the compressor $|\beta_1 \ldots
\beta_{\lambda(n)}\rangle$ are taken as the {\it compressed version}
of its input, and the remaining $n-\lambda(n)$ qubits are discarded.
A {\it decompressor B} is a unitary transformation that maps $n$-qubit
strings to $n$-qubit strings.  It is to be understood that the first
$\lambda(n)$ qubits input to the decompressor are $|\beta_1 \ldots
\beta_{\lambda(n)}\rangle$, the compressed version of some sequence of
$n$ qubits, and the remaining $n-\lambda(n)$ qubits are all
$|0\rangle$.  An {\it $n$-to-$\lambda(n)$ quantum compression scheme}
is a compressor/decompressor pair $(A,B)$.  As in the classical case,
the goal is to achieve as high a compression rate (i.e. as small a
$\lambda(n)$) as possible, while permitting the original message to be
recovered from its compressed version, with high probability.

Assume that the compressor knows (i.e. can be a function of) the
underlying ensemble $(\S,p)$, but has no explicit knowledge
about the specific random selections made (interestingly, compressors
exist that know even less than $(\S,p)$; more about this later).  In
the classical case, the compressor obtains complete information about
the bits to be compressed, but complete information cannot generally
be obtained from a qubit.  If the possible qubit states in $\S$ are
not mutually orthogonal then any observation of such a qubit will only
yield partial information about its state, and can irretrievably
change this state.  Due to this, one might expect to be able to
achieve less in the quantum scenario than with classical compression
schemes --- in fact, the opposite is true.

Let us measure the quality of an $n$-to-$\lambda(n)$ compression scheme
$(A,B)$ with respect to a source distribution $p$ in terms of its
{\it fidelity}, defined as follows.
Consider the following experiment.
Let the sequence $|\alpha_1\rangle \ldots |\alpha_n\rangle$ be sampled
independently from $(\S,p)$.
Transform $|\alpha_1\rangle \ldots |\alpha_n\rangle$ according to the
compressor $A$ and let $|\beta_1 \ldots \beta_{\lambda(n)}\rangle$
be the compressed version.
Next, transform
$|\beta_1 \ldots \beta_{\lambda(n)}\rangle
|0 \ldots 0\rangle$ according to the decompressor $B$
and let
$|\alpha^{\prime}_1 \ldots \alpha^{\prime}_n\rangle$
be the output.
Finally, measure $|\alpha^{\prime}_1 \ldots \alpha^{\prime}_n\rangle$
with respect to a basis containing $|\alpha_1 \ldots \alpha_n\rangle$.
The {\it fidelity} is the probability
$|\langle\alpha^{\prime}_1 \ldots \alpha^{\prime}_n|
\alpha_1 \ldots \alpha_n\rangle|^2$ that this measurement results
in $|\alpha_1 \ldots \alpha_n\rangle$.

Note that the fidelity is with respect to two sources of randomness:
(a) the random choices in the original generation of
$|\alpha_1 \ldots \alpha_n\rangle$; and (b) the randomness that results
from performing a measurement of the state
$|\alpha^{\prime}_1 \ldots \alpha^{\prime}_n\rangle$.
Roughly speaking, the fidelity can be high if for ``most'' choices
in (a), $|\alpha^{\prime}_1 \ldots \alpha^{\prime}_n\rangle$ is
``close to'' $|\alpha_1 \ldots \alpha_n\rangle$.


The ensemble $(\S,p)$ represents a {\em mixed state}, which has
{\it density matrix} $\rho$, defined as
$$\rho = \sum_{i=1}^m p_i |\Psi_i\rangle\langle\Psi_i|.$$
The {\em von Neumann entropy} corresponding to $(\S,p)$ is defined in terms
of the density matrix $\rho$ as $\Hvn(\rho) = - \mbox{Tr}(\rho \log \rho)$.
In general, $\Hvn(\rho) \le \Hs(p)$, with equality occurring if and only if
the states in $\S$ are mutually orthogonal.

Roughly speaking, Schumacher's theorem \cite{Schum1} states that
$n \Hvn(\rho)$ is asymptotically the maximum compression attainable for
$n$ qubits resulting from a source with density matrix $\rho$.
More precisely, let $(\S,p)$ be any ensemble of qubits,
and $\rho$ be the corresponding density matrix.
Then, for all $\e, \d > 0$, for sufficiently large $n$ and
$\lambda(n) \ge n (\Hvn(\rho) + \d)$,
there exists an $n$-to-$\lambda(n)$ quantum compression scheme for
$(\S,p)$ with fidelity greater than $1 - \e$.
Moreover, for all $\e, \d > 0$,
for sufficiently large $n$, if $\lambda(n) \le n (\Hvn(\rho) - \d)$ then
every $n$-to-$\lambda(n)$ quantum compression scheme has fidelity less than
$\e$.

It should be noted that the above bounds are robust in the sense that
they do not change when a number of technical variations are made in
the scenario.  For example, the $n$-to-$\lambda(n)$ compression
schemes that attain fidelity greater than $1 - \e$ can restricted to
being highly ``oblivious'' in that they depend only on knowing a basis for
which the density matrix is diagonal, with nonincreasing values along
the diagonal.  Also, \cite{SJ} even if the compressor is supplied with
complete information about the state of the source string $|\alpha_1
\ldots \alpha_m \rangle$ that it receives, $\e$ still bounds the
fidelity attainable if $\lambda(n) \le n (\Hvn(\rho) - \d)$.

The proof of Schumacher's Theorem is based on the existence of a
``typical subspace" $\Lambda$ of the Hilbert space of $n$ qubits,
which has the property that, with high probability, a sample of
$|\alpha_1,\ldots,\alpha_n\rangle$ has almost unit projection onto
$\Lambda$.  It has been shown\cite{Schum1,SJ} that the dimension of
$\Lambda$ is $2^{n\Hvn(\rho)}$; thus, the operation that the
compressor should perform involves ``transposing" the subspace
$\Lambda$ into the Hilbert space of a smaller block of $n\Hvn(\rho)$
qubits.

Bennett\cite{BPT} gives a more explicit procedure for accomplishing
this ``transposition", which we illustrate with an example.  Suppose
that $\S = \{|\Psi_1\rangle,|\Psi_2\rangle\}$, where
$|\Psi_1\rangle=|0\rangle$ and
$|\Psi_2\rangle=\frac{1}{\sqrt{2}}|0\rangle +
\frac{1}{\sqrt{2}}|1\rangle$, and $p=(p_1,p_2)$, where
$p_1=p_2=\frac{1}{2}$.  The density matrix corresponding to $(\S,p)$
is $\rho=\frac{1}{2}|0\rangle\langle 0|+\frac{1}{2}
(\frac{1}{\sqrt{2}}|0\rangle + \frac{1}{\sqrt{2}}|1\rangle)
(\frac{1}{\sqrt{2}}\langle 0|+\frac{1}{\sqrt{2}}\langle 1|)$, or, in
2$\times$2 matrix form,
\begin{equation}
\rho=\left (\begin{array}{rr}\frac{3}{4}&\frac{1}{4}\\ \frac{1}{4}&
\frac{1}{4}\end{array}\right)\mbox{ in the basis }\begin{array}
{r}|0\rangle\\|1\rangle\end{array}.\label{rho1}
\end{equation}
It is always possible to go to a basis in which the density matrix
is diagonal:
\begin{equation}
\begin{array}{ll}\rho^{\prime}&
{=\left (\begin{array}{cc}\lambda_{max}&0\\0&\lambda_
{min}\end{array}\right )}\\
&{=\left (\begin{array}{cc}\frac{3}{4}+\frac{1}{4}\tan\frac{\pi}{8}&0\\
0&\frac{1}{4}-\frac{1}{4}\tan\frac{\pi}{8}\end{array}\right )\mbox{ in
the basis }
\begin{array}{l}
|0^{\prime}\rangle =\cos\frac{\pi}{8}|0\rangle+\sin\frac{\pi}{8}
|1\rangle\\
|1^{\prime}\rangle =-\sin\frac{\pi}{8}|0\rangle+\cos\frac{\pi}{8}|1\rangle
\end{array}.}\end{array}\label{rho2}
\end{equation}
Both of the states $|\Psi_1\rangle$ and $|\Psi_2\rangle$ have large
overlap on the basis state $|0^{\prime}\rangle$
($|\langle\Psi_i|0^{\prime}\rangle|=\cos\frac{\pi}{8}$), and small
overlap on the orthogonal basis state $|1^{\prime}\rangle$
($|\langle\Psi_i|1^{\prime}\rangle|=\sin\frac{\pi}{8}$).  This
observation leads to a way of compressing strings of signal states.
Consider all $n$-qubit strings possible from the states in $\S$.
These strings can all be expressed with respect to the basis
consisting of $|x_{n-1} \ldots x_0\rangle = |x_{n-1}\rangle \ldots
|x_0\rangle$, where $x_{n-1}, \ldots, x_0 \in
\{|0^{\prime}\rangle,|1^{\prime}\rangle\}$.  Each such $|x_{n-1}
\ldots x_0\rangle$ can be interpreted as an $n$-bit binary number,
and, thus, can be denoted as $|x\rangle$, for $x \in
\{0,\ldots,2^n-1\}$.  Now, the overlap of $|x\rangle$ with the states
in $\S^n$ is $|\langle x | \S^n \rangle| =
\cos^m\frac{\pi}{8}\sin^{n-m}\frac{\pi}{8}$, where $m$ is the number
of 0's in the binary representation of $x$.  Because this overlap
diminishes exponentially with $n-m$, basis states with large numbers
of 1's are relatively unimportant for describing any string
$|\alpha_1,\ldots,\alpha_n\rangle$; the Hilbert space can thus be
truncated to the typical subspace $\Lambda$ consisting of all states
$|x\rangle$ in which the binary number $x$ contains a proportion of
1's less than $\Hvn(\rho) < 0.601$.

Thus the ``transposition" which the coder must do consists of
mapping this $\Lambda$ subspace for $n$ qubits into the states
spanned by less than $0.601n$ of those qubits.

We must accomplish this by a unitary transformation applied to the
original states of the $n$ qubits.  In the basis
$|0\rangle,\ldots,|2^n-1\rangle$, this transformation must map qubit
strings with the smallest number of 1's in succession into qubit
binary strings with the smallest numerical value.  This is a classical
combinatorial calculation, ``classical'' in the sense that definite
binary-number states are mapped to other definite binary-number
states; however, it is essential that the computation be performed
quantum mechanically, since the computation must preserve the
superpositions of these basis states.  This means that the
combinatorial computation must be performed using reversible,
quantum-coherent elementary operations.

The principal object of this paper is to derive the quantum
computation which is needed to do this Schumacher coding.  In
Sec. \ref{sec:two} we derive the analytical formula for the sorting
calculation required for the coding.  Sec. \ref{sec:prog} constructs
the quantum program for performing this calculation:
Sec. \ref{sec:attempt} illustrates a first attempt at this coding
exercise; Sec. \ref{sec:rev} discusses the way in which the
calculation is to be properly made reversible; and
Sec. \ref{sec:final}, which contains the essential result of the
paper, gives the final quantum program for Schumacher coding.
Sec. \ref{sec:bit} gives, in the same programming notation developed
in the earlier sections, the bit-level routines needed for performing
the steps in the high-level program.  Appendix \ref{ap:pebble}
discusses how these bit-level routines may be made highly
space-efficient, with only a modest increase in running time (these
latter routines result in a smaller time-space product, which may be
desirable \cite{Unruh,Palma}).  Appendix \ref{ap:phase} provides other
ways of economizing in the bit-level implementation of these codes, by
using some of the phase freedom coming from the quantum-mechanical
nature of the computation.

\section{Combinatorial expression for Schumacher coding}
\label{sec:two}

As Bennett\cite{BPT} has described, a specific realization of the
unitary transformation performing the Schumacher coding function on a
set of identical qubits consists of a sorting computation in which the
states $|0\rangle,\ldots,|2^n-1\rangle$ are given a lexicographical
ordering according to how many 1's are in their binary expansion.  So,
$|0\rangle$ is mapped to itself, all the states containing exactly one
1 and $n-1$ 0's are mapped to the states between $|1\rangle$ and
$|n\rangle$, all the states with exactly two 1's and $n-2$ 0's are
mapped to the states between $|n+1\rangle$ and $|n+n(n-1)/2\rangle$,
and, in general, all the states with exactly $m$ 1's and $n-m$ 0's are
mapped to the states between
\begin{equation}
\textstyle{\left|\sum_{i=0}^{m-1}{n\choose i}\right\rangle}\label{ineq1}
\end{equation}
and
\begin{equation}
\textstyle{\left|\sum_{i=0}^{m}{n\choose i}-1\right\rangle}\label{ineq2}
\end{equation}
inclusive.  The Schumacher function does not require any particular
ordering of the states within each of these blocks, except that the
mapping must be 1-to-1 (i.e., a bijection); but, it turns out to be
convenient to preserve lexicographical ordering within each block.
Defining the index number within each block as $I[x,n,m]$, the total
Schumacher function for string $x$ (with $n$ bits and $m$ 1's) is
\begin{equation}
y=\sum_{i=0}^{m-1}{n\choose i}+I[x,n,m].\label{breakdown}
\end{equation}
The index number $I$ obeys a recursive relationship which we now
derive.  Considering the possible binary-number strings representing
the input state $x$, any string whose first 1 occurs in the
$p+1^{\mbox{\scriptsize st}}$ place (i.e., whose first $p$ bits are 0)
must have a higher index number than all strings in which the first
$p+1$ places are 0.  There are exactly ${{n-p+1}\choose{m}}$ such
strings.  This means that for the particular input string
\begin{equation}
x=\overbrace{00\ldots 00}^{p\ 0's}1\overbrace{000\ldots 00}^{n-m-p\ 0's}
\overbrace{111\ldots 11}^{\;\;m-1\ 1's},\label{stringdec}
\end{equation}
the index number $I[x,n,m]={{n-p-1}\choose{m}}$.  This result permits the
index number of the more complex string
\begin{equation}
x=\overbrace{00\ldots 00}^{p\ 0's}1\!\!\!\!\!\!\!\!\!
\underbrace{x^{\prime}}_{\mbox{\scriptsize $n-p-1$ bits}}\label{substr}
\end{equation}
to be expressed recursively:
\begin{equation}
I[x,n,m]={{n-p-1}\choose{m}}+I[x^{\prime},n-p-1,m-1].\label{recur}
\end{equation}
It is probably easiest to understand Eq. (\ref{recur}) by writing out
an example:
\begin{eqnarray}
I[0010011011,10,5]=\mbox{${{10-2-1}\choose 5}$}+I[0011011,7,4]
\;\;\;\;\;\;\;\;\;\;\;\;\;\;\;\;\;\;\;\;\;\;\;\;\;\;\;\;\;\;\;\;
\;\;\;\;\;\;\;\;\;\;\nonumber\\
\Downarrow\;\;\;\;\;\;\;\;\;\;\;\;\;\;\;\;\;\;\;\;\;\;
\;\;\;\;\;\;\;\;\;\;\;\;\;\;\;\;\;\;\;\;\;\;\;\;\;\;\;\;\;\nonumber\\
\mbox{${{7-2-1}\choose 4}$}+I[1011,4,3]\;\;\;\;\;\;\;\;\;\;\;\;\;\;\;\;
\;\;\;\;\;\;\;\;\;\;\label{examp}\\
\Downarrow\;\;\;\;\;\;\;\;\;\;\;\;\;\;\;\;\;\;\;\;\;\;\;\;\;\;\;\;
\nonumber\\
\mbox{${{4-0-1}\choose 3}$}+I[011,3,2]\;\;\;\;\nonumber\\
\Downarrow\;\;\;\;\;\;\;\;\;\;\;\;\nonumber\\
0.\;\;\;\;\;\;\;\;\;\;\;\;\nonumber
\end{eqnarray}
As this illustrates, the recursion of Eq. (\ref{recur}) may be
iterated to produce an expression for $I$ for a general input string $x$:
\begin{equation}
I[x,n,m]=\sum_{i=1}^{n-1}x_{n-i}{{n-i}\choose{\sum_{k=i}^n
x_{n-k}}}.\label{isum}
\end{equation}
Here the notation $x_p$ denotes the value of the $p^{\mbox{\scriptsize
th}}$ bit of the string $x$.  Combining Eq. (\ref{isum}) with
Eq. (\ref{breakdown}) yields the final expression for the Schumacher
coding function:
\begin{equation}
y=\sum_{i=0}^{\sum_{k=0}^{n-1}x_k-1}
{n\choose i}+
\sum_{j=1}^{n-1}x_j{j\choose{\sum_{k=0}^j x_k}}.\label{scf}
\end{equation}
In this equation, binary coefficients outside their natural range
(e.g., ${n\choose{n+1}}$) are understood to be zero.

\section{High-Level quantum program for Schumacher coding}
\label{sec:prog}
\subsection{first attempts}
\label{sec:attempt}

It is now our object to translate Eq. (\ref{scf}) into a sequence of
elementary quantum-mechanical manipulations.  We proceed to do this by
writing out the calculation in a high-level ``pseudocode'' \cite{Brau}
which, when ``compiled'', would permit the operation to be performed by
a sequence of elementary spectroscopic manipulations such as
two-bit XOR's (or controlled-NOT's), along with one-bit
rotations\cite{G9}.  Rather than building up the rules of this
pseudocode axiomatically, we will proceed in an intuitive fashion.
The principal constraint which the coded calculation must obey is that
it be done {\em reversibly}.  Instead of going into a discourse about this,
let us present the first try at coding Eq. (\ref{scf}) (not a perfectly
successful one, in fact):

\ni {\bf Program FIRST\_TRY \nl
\ii quantum registers:} \nl
\ii $X$ : $n$-bit register \nl
\ii $Y$ : $n$-bit arithmetic register (initialized to 0) \nl
\ii $S$ : $\lceil\log n\rceil$-bit arithmetic register (initialized to 0) \nl
\nl{\bf
\ii if $X_0 = 1$ then $S \- S + 1$ \nl
\ii for $j=1$ to $n-1$ do \nl
\ii \ii if $X_j = 1$ then $S \- S + 1$ \nl
\ii \ii for $m = 0$ to $j+1$ do \nl
\ii \ii \ii if $X_j = 1$ and $S = m$ then
$Y \- Y + {j \choose m}$ \nl
\ii for $i=0$ to $n-1$ do \nl
\ii \ii if $i+1 \le S$ then
$Y \- Y + {n \choose i}$}\ee

{\bf FIRST\_TRY} is not incorrect, but it is incomplete, in ways which
we will repair by stages below.  Here are some rules of this
programming: All the quantum-mechanical registers are in capital
letters.  In {\bf FIRST\_TRY}, these are $X$ (which is initialized
with an input state $x$, or a quantum superposition of such input
states), $Y$ (which is initialized to 0, and whose final value is the
output state $y$ or their quantum superpositions), and $S$ (a small
work register, also initialized to 0).  The notation $X_i$ indicates
the $i^{\mbox{\scriptsize th}}$ bit of $X$.  Note that $Y$ and $S$ are
given the data type ``arithmetic'', indicating that ordinary integer
addition and subtraction are allowed with them.  Only bitwise
manipulations are performed on $X$.  (In the {\bf FINAL\_SCHUMACHER}
program, both bitwise and arithmetic manipulations will be performed
on the same registers.)

All other lower-case variables in the program always have definite
values and can (and should) be implemented using classical bits.  Only
the quantum registers need to be explicitly treated reversibly.  So,
the binomial coefficients ${j \choose m}$ can be precomputed or
evaluated by any means, reversible or not, in the implementation of
the quantum computation.

In a reversible program statement, the input can always be deduced
from the output.  So, for example, the statement

{\bf\ni \ii if $X_0 = 1$ then $S \- S + 1$}

\ni is reversible, because the input could be deduced by the
``time-reverse'' of this statement,

{\bf\ni \ii if $X_0 = 1$ then $S \- S - 1$}

\ni An irreversible program statement would be

{\bf\ni \ii if $X_0 = 1$ then $S \- 1$}

\ni since the prior value of $S$ cannot be deduced.  As it happens,
this statement would function correctly in {\bf FIRST\_TRY} because
$S$ actually is equal to 0 at this first executable statement of the
program.  However, we will enforce a rule that the only irreversible
statements permitted that involve quantum variables will be the
``initialized" designations present in the declaration statements.  In
later programs we will introduce a ``finalized" designation, which
will merely serve as a reminder that certain variables will always end
the program with a particular value if the program runs correctly.
This designation will be an important one in constructing reversible
code.  It is also a reminder that physically, the finalization can
serve as a useful check that no error has occurred\cite{Chuang}; a
quantum measurement of this register at the end of the running of
program should always find the register in the finalized value.

One further comment about the program statement

{\bf\ni \ii if $X_0 = 1$ then $S \- S + 1$}.

\ni If $S$ were a one-bit variable, this statement would just be a
quantum XOR or controlled NOT, in which the value of $S$ is inverted
conditional on the value of $X_0$.  In {\bf FIRST\_TRY}, $S$ is a
multibit register, in fact it must have about $\log_2n$ bits.
Implementation of these multi-bit functions in terms of primitive
operations involving no more than three bits is straightforward, and
is presented in Sec. \ref{sec:bit} and in Ref. \cite{Vedral,Beckm}.
Using quantum gates, all the three-bit primitives may be reduced to
sequences of two-bit operations\cite{G9}.

A few more points about {\bf FIRST\_TRY} are in order.  Given the
constraints of reversibility, it is a relatively straightforward
transcription of Eq. (\ref{scf}).  The first {\bf for} loop (indexed
by $j$) implements the second term of Eq. (\ref{scf}); this is
efficient because the partial sum in the binomial coefficient can be
accumulated in $S$ one term at a time, and then the completed sum can
be used as the upper limit of the first term of (\ref{scf}), which is
implemented in the second {\bf for} loop.  This inner $m$ loop could be
replaced by the single statement

{\bf\ni \ii if $X_j = 1$ then $Y \- Y + {j \choose S}$},

\ni but this would require a reversible calculation of the binomial
function; we have chosen to make this binomial-coefficient calculation
classical by writing out the loop as shown.  One might also be tempted
to modify the inner loop as follows:

{\bf\ni
\ii if $X_j = 1$ then\nl
\ii \ii for $m = 0$ to $j+1$ do \nl
\ii \ii \ii if $S = m$ then $Y \- Y + {j \choose m}$}

\ni While moving the {\bf if} statement from the loop is superficially
more efficient, it turns out that, when these statements are
re-expressed in terms of primitive operations, the {\bf if} must be
carried down to the lowest level in any case; so, we prefer a syntax
in which such conditionals are explicitly shown at the lowest level.

\subsection{Reversibility considerations}
\label{sec:rev}

Now, what is the overall effect of {\bf FIRST\_TRY}, and why is it
inadequate for performing the Schumacher function?  Let $f : \{0,1\}^n
\rightarrow \{0,1\}^n$ denote the Schumacher function for $n$-bit
binary strings.  If the total input state is expressed as the ket
\begin{equation}
|X,Y,S\rangle = |x,0,0\rangle,\label{initialshoe}
\end{equation}
then the complete final state is
\begin{equation}
|X,Y,S\rangle = |x,f(x),s\rangle,\label{badshoe}
\end{equation}
(where $s$ is the number of 1's in $x$).  But, the correct Schumacher
function must have a final state of the form
\begin{equation}
|X,Y,S\rangle = |0,f(x),0\rangle.\label{goodshoe}
\end{equation}
That is, the input $x$ should be erased and the work register $S$
should be reset to its initial value of 0.  This is possible to
accomplish reversibly because the Schumacher function is bijective, so
that no record of the initial state, or of the state of the work bits,
needs to be retained at the end; they are completely deducible from
the output.  In fact, the correct operation of the Schumacher function
{\em requires} that the output be of the form (\ref{goodshoe}); if it
is of the form of (\ref{badshoe}), then the final state is
``entangled'' with the initial state, which means that output states
cannot be placed in the desired superpositions of states.  Thus, the
net result of the Schumacher function should be confined to the input
data register only; this condition is obtainable from
Eq. (\ref{goodshoe}) if the final output state is swapped so that the
state vector becomes
\begin{equation}
|X,Y,S\rangle = |f(x),0,0\rangle.\label{swapshoe}
\end{equation}
Thus the Schumacher function is applied, ``in-place'', to the first
$n$ qubits, while the remaining $n+\log_2n$ bits return to their
original states, and may all be viewed simply as work space for the
computation.  We will see later that the ``output" register $Y$ can
actually be removed entirely by using some clever programming.  Some
other workspace, not displayed explicitly in (\ref{swapshoe}), appears
to be necessary to do the bit-level manipulations in the Schumacher
function (see Sec. \ref{sec:bit}); Appendix \ref{ap:pebble} shows that
the size of this extra workspace does not have to exceed about
$2\sqrt{n}$ bits.

These considerations have arisen previously in the context of
reversible programming\cite{Bennett}, but the rationale for
constructing a function in the ``fully-reversible'' manner as
specified by the output state (\ref{swapshoe}) is somewhat different
than in the classical context.  In traditional reversible programming
the object is to avoid the small energy cost involved in irreversible
erasure of any of the working bits in the computer.  If such an
erasure is performed, the {\em result} of the computation will still
be correct, even though the desired goal of expending no energy is not
achieved.  But in quantum computation, irreversible erasure of the
state of register $X$ in Eq. (\ref{badshoe}) actually causes register
$Y$ to be in the {\em wrong} quantum state, in so far that, if the
initial $X$ was in a superposition of computational states, the final
state of $Y$ will be a mixed quantum state, rather than the intended,
pure superposition state.  Thus, the consequences of irreversibility
are more serious than in conventional reversible computation.

A method for designing a calculation to arrive at the desired final
states (\ref{goodshoe}) or (\ref{swapshoe}), as already worked out in
the earlier literature\cite{Bennett}, requires two steps: 1) zero out
$S$ and any other workspaces used by the program, and 2) explictly
implement the {\em inverse} of the Schumacher function
Eq. (\ref{scf}).  This can be accomplished by a program that, on input
state
\begin{equation}
|X,Y,S\rangle = |x,y,0\rangle,\label{initialinvshoe}
\end{equation}
produces the final state
\begin{equation}
|X,Y,S\rangle = |x \xor f^{-1}(y),y,0\rangle.\label{invshoe}
\end{equation}
Note that applying such a transformation to the state
\begin{equation}
|X,Y,S\rangle = |x,f(x),0\rangle\label{partfix}
\end{equation}
yields the required state
\begin{equation}
|X,Y,S\rangle
= |0,f(x),0\rangle.\label{inv3shoe}
\end{equation}

Eq. (\ref{invshoe}) is not implemented simply by running {\bf
FIRST\_TRY} in reverse; indeed, the inverse function can have very
different and much greater complexity than the function
itself\cite{Chau1way}.  Fortunately, in this case, as we will see in a
moment, the inverse Schumacher function is also relatively easy to
implement.

Step (1) above, zeroing out $S$, is readily performed by adding code
to the end of {\bf FIRST\_TRY} to simply subtract away the bits which
have been added to $S$:

{\bf
\ni
\ii for $j = 0$ to $n-1$ do \nl
\ii \ii if $X_j=1$ then $S \- S-1$}

\ni This code, added to the end of {\bf FIRST\_TRY}, produces the output
state (\ref{partfix}).

Step (2) above, implementing the inverse function Eq.
(\ref{invshoe}), requires a new algorithm.  We have not found any way
to write the inverse Schumacher coding function as a formula as in Eq.
(\ref{scf}).  Nevertheless, a straightforward algorithm can be deduced
from the following two inequalities.  The first is obtained by
combining the information from Eqs. (\ref{ineq1}) and (\ref{ineq2}):
\begin{equation}
\sum_{i=0}^{m-1}{n\choose i}\le y < \sum_{i=0}^m{n\choose i},
\label{bound1}
\end{equation}
where
\begin{equation}
m=\sum_{k=0}^{n-1}x_k
\end{equation}
is the number of 1's in the binary string $x$.  We will be able to
write simple pseudocode to compute $m$ (a.k.a. $S$).  This result can
then be used to compute $I[x,n,m]$ using Eq. (\ref{breakdown}).
$I[x,n,m]$ satisfies an inequality which is a simple consequence of
Eq. (\ref{recur}) and the discussion preceding it:
\begin{equation}
{{n-p-1}\choose{m}}\le I[x,n,m] < {{n-p}\choose{m}}.\label{bound2}
\end{equation}
By finding the $p$ which satisfies this equation, we determine that the
leading $p$ bits of $x$ are zeros, and the next bit is a 1 (i.e.,
$x_j=0$, $n-1-p < j \le n-1$, $x_{n-p-1}=1$).  The index of the
remaining substring can be determined from Eq. (\ref{recur}), and thus all
the bits of $x$ may be calculated recursively.

\subsection{Deriving the final program}
\label{sec:final}

Now we will transform our procedure into reversible code.  As the last
section makes clear, a necessary step for doing this will be to code
the inverse of the Schumacher function.  In the spirit of {\bf
FIRST\_TRY}, we will not worry at first about the final state of the
work registers as prescribed in Eq. (\ref{invshoe}); we will initially
just try to code correctly the inverse function itself.  We will find
that reversibility will, in this case, fall out naturally from a
simple modification of our first-cut program.

\ni {\bf Program TRY\_INVERSE \nl
\ii quantum registers:} \nl
\ii $X$ : $n$-bit register (initialized to 0) \nl
\ii $Y$ : $n$-bit signed arithmetic register (finalized to 0) \nl
\ii $S$ : $\lceil\log n\rceil$-bit register (initialized and finalized
to 0) \nl
\nl{\bf
\ii for $m=0$ to $n$ do \nl
\ii \ii $Y \- Y - {n \choose m}$ \nl
\ii \ii if $Y \ge 0$ then $S \- S + 1$ \nl
\ii for $m=0$ to $n$ do \nl
\ii \ii if $S \ge m$ then $Y \- Y + {n \choose m}$ \nl
\ii for $p=0$ to $n-1$ do \nl
\ii \ii for $i=0$ to $n-p$ do \nl
\ii \ii \ii if $S=i$ and $Y \ge {{n-p-1}\choose i}$ then
$X_{n-p-1} \- X_{n-p-1} \xor 1$\nl
\ii \ii \ii if $S=i$ and $X_{n-p-1}=1$ then $Y \- Y-{{n-p-1}\choose i}$ \nl
\ii \ii if $X_{n-p-1}=1$ then $S \- S-1$}

In this code, the $m$-loop does the job of finding the $m$ for which
Eq. (\ref{bound1}) is satisfied, and putting the result in the quantum
register $S$.  As a byproduct of this work, it subtracts away the
first term of Eq. (\ref{scf}) from $y$, leaving in $Y$ the value of
the index $I[x,n,m]$.  Actually, the $m$-loop continues to subtract
binomial coefficients from $Y$ after it is supposed to; this is why
$Y$ is indicated to be a ``signed'' register, which can be handled by
doing ordinary arithmetic in a register with one extra bit (see
\cite{Vedral}).  This approach has the benefit that testing that $Y$
is non-negative only requires the examination of one bit --- see the
first part of Sec. \ref{sec:bit}.  We might be tempted to avoid
negative numbers by terminating the loop at the right moment, viz:

{\bf
\ni
\ii for $m=0$ to $n$ do \nl
\ii \ii if $Y < {n \choose m}$ then exit for-loop}\nl
\ii \ii $\vdots$

\ni But such an {\bf exit for-loop} statement is not reversible.
There appears to be no alternative to letting the first loop go to its
maximum possible upper limit, which is $n$, and then repairing the
damage done by adding back the correct binomial coefficients in the
second loop.  Finally, at the end of the second loop, $Y$ has the
desired value of $I[x,n,m]$, and $S$ has the value of $m$.

Then the third ($p$) loop of {\bf TRY\_INVERSE} does the iterative
decomposition of the index $I[x,n,m]$.  For every possible value of
the leading number of zeros $p$ (recall Eq. (\ref{stringdec})), {\bf
TRY\_INVERSE} checks to see if the inequality Eq. (\ref{bound2}) is
satisfied; if it is, then the program negates one bit of the $X$
register.  Then the second {\bf if} statement decrements $Y$ by the
combinatorial coefficient in Eq. (\ref{recur}), so that it always
contains the index of the next substring.  The process continues until
the index is reduced to zero.  Also, $S$ is decremented so that it
always contains the current value of the number of 1's in the
substring of Eq. (\ref{substr}).  Note that, as in {\bf FIRST\_TRY},
an inner loop (indexed by $i$) is introduced to avoid the need for
reversible calculation of binomial coefficients like ${{n-p-1} \choose
S}$.

We now evaluate what state {\bf TRY\_INVERSE} has left the registers
$Y$ and $S$ in. In fact, a very desirable thing has ``accidentally''
occurred!  We find that, on input state
\begin{equation}
|X,Y,S\rangle = |0,y,0\rangle,\label{accident1}
\end{equation}
{\bf TRY\_INVERSE} produces the final state
\begin{equation}
|X,Y,S\rangle = |f^{-1}(y),0,0\rangle.\label{accident2}
\end{equation}

Thus, with a final transposing of the $X$ and $Y$ registers, we obtain a
program that implements the {\it inverse} of Eq. (\ref{swapshoe}), so
the calculation has been successfully done in-place, with the registers
$Y$ and $S$ remaining in their initial state, having served only as
``catalysts'' for the calculation.

In fact, we can do even better; by a small modification of {\bf
TRY\_INVERSE}, the $Y$ register can be eliminated entirely.  This can
be done by noting that, during the course of an execution of {\bf
TRY\_INVERSE}, the decrementing of $Y$ sets each of its high-order
bits to zero in succession, and, at the same time, the values of $X$
are built up starting with the high-order bits and working down.
Thus, the high-order bits of $Y$ can be re-used to hold the results of
the final calculation.  It can be shown that these high-order bits are
always cleared out soon enough that they can be used for the final
answer; this is done by showing that in {\bf TRY\_INVERSE}, the same
bits of $X$ and $Y$ are never simultaneously 1.  Thus, with one small
modification, {\bf TRY\_INVERSE} can be turned into our final program
for the inverse of the Schumacher coding function:

\ni {\bf Program FINAL\_SCHUMACHER\_INVERSE \nl
\ii quantum registers:}\nl
\ii $X$ : $n$-bit signed arithmetic register \nl
\ii $S$ : $\lceil\log n\rceil$-bit arithmetic register
(initialized and finalized to 0) \nl
\nl{\bf
\ii for $m=0$ to $n$ do \nl
\ii \ii $X \- X - {n \choose m}$ \nl
\ii \ii if $X \ge 0$ then $S \- S + 1$ \nl
\ii for $m=0$ to $n$ do \nl
\ii \ii if $S \ge m$ then $X \- X + {n \choose m}$ \nl
\ii for $p=0$ to $n-1$ do \nl
\ii \ii for $i=0$ to $n-p$ do \nl
\ii \ii \ii if $S=i$ and
$\mbox{TRUNC}_{n-p-1}(X) \ge {{n-p-1}\choose i}$ then
$X_{n-p-1} \- X_{n-p-1} \xor 1$\nl
\ii \ii \ii if $S=i$ and $X_{n-p-1}=1$ then $X \- X-{{n-p-1}\choose i}$ \nl
\ii \ii if $X_{n-p-1}=1$ then $S \- S-1$}

The only substantial item which has been added here is the function
{\bf $\mbox{TRUNC}_j$}.  Invocation of {\bf $\mbox{TRUNC}_j({\sl X})$}
simply says that only the $j$ least significant bits of the quantum
register $X$ (i.e., bit 0 to bit $j-1$) should be taken account of in
the ``$\ge$'' comparison.  This is necessary because the high-order
bits are being used to store the final answer.  In the final pass
through the $p$ loop, the occurrence of the zero index in {\bf
$\mbox{TRUNC}_0({\sl X})$} indicates that the comparison should not be
performed at all.

For completeness, we now record the final code for the Schumacher
coding function itself.  Since {\bf FINAL\_SCHUMACHER\_INVERSE} is
done ``in-place'', the direct function is literally just the
time-reverse:

\ni {\bf Program FINAL\_SCHUMACHER \nl
\ii quantum registers:}\nl
\ii $X$ : $n$-bit signed arithmetic register \nl
\ii $S$ : $\lceil\log n\rceil$-bit arithmetic register
(initialized and finalized to 0) \nl
\nl{\bf
\ii for $p=n-1$ down to $0$ do \nl
\ii \ii if $X_{n-p-1}=1$ then $S \- S+1$\nl
\ii \ii for $i=n-p$ down to $0$ do \nl
\ii \ii \ii if $S=i$ and $X_{n-p-1}=1$ then $X \- X+{{n-p-1}\choose i}$ \nl
\ii \ii \ii if $S=i$ and
$\mbox{TRUNC}_{n-p-1}(X) \ge {{n-p-1}\choose i}$ then
$X_{n-p-1} \- X_{n-p-1} \xor 1$\nl
\ii for $m=n$ down to $0$ do \nl
\ii \ii if $S \ge m$ then $X \- X - {n \choose m}$ \nl
\ii for $m=n$ down to $0$ do \nl
\ii \ii if $X \ge 0$ then $S \- S - 1$ \nl
\ii \ii $X \- X + {n \choose m}$}

\section{Bit-level quantum program for Schumacher coding}
\label{sec:bit}

In this section, we explain how the statements in programs {\bf
FINAL\_SCHUMACHER} and {\bf FINAL\_SCHUMACHER\_INVERSE} can be
implemented by a gate-array with fundamental bit-level operations.
These fundamental operations are essentially Toffoli
gates\cite{Toffoli80}.  The Toffoli gate that negates bit $B$ iff bits
$C$ and $D$ are both 1 (and doesn't change the values of $C$ and $D$)
is denoted as

\ni\ii $B \- B \xor (C \wedge D)$.

\ni In \cite{G9} it is shown that such an operation can be simulated
in terms of eight one-bit operations and eight XOR operations (which
are of the form $B \- B \xor C$).  For convenience, we expand our
repertoire of allowable basic operations to include

\ni
\ii $B \- B \xor 1$ \nl
\ii $B \- B \xor C$ \nl
\ii $B \- B \xor \overline{C}$ \nl
\ii $B \- B \xor (\overline{C} \wedge D)$ \nl
\ii $B \- B \xor (\overline{C} \wedge \overline{D})$ \nl
\ii $B \- B \xor (C \vee D)$.

\ni As with Toffoli gates, each of these gates can be simulated by at most
eight one-bit operations and eight XOR operations.  In many cases a quantum
phase freedom can be used to simulate these in fewer one- and two-bit gates
(see Appendix \ref{ap:phase}).

The first step to converting the programs into gate-arrays is to
``unravel'' the {\bf for} loops.  Since the ranges of these loops are
all fixed prior to any computation, this is straightforward.  Next, we
note that (once the {\bf for} loops have been unravelled) there are
essentially five types of program statements:
\begin{enumerate}
\item $X \- X + k$
\item {\bf if $B$ then $X \- X + k$}
\item {\bf if $Y > l$ then $X \- X + k$}
\item {\bf if $Y = l$ and $B$ then $X \- X + k$}
\item {\bf if $Y = l$ and $Z > k$ then $B \- B \xor 1$}
\end{enumerate}
(where $B$ is a bit, $X$, $Y$, $Z$ are signed arithmetic registers,
and $k$, $l$ are signed integers).

Also, there are {\em a priori} upper bounds on the ranges of the
arithmetic registers (and thus on the number of bits required to
specify them).  An arithmetic register whose range of values is known
to be an integer within $[0,2^n)$ can be naturally represented by $n$
bits and arithmetic operations on it can be simulated by reversibly
performing them modulo $2^n$.  Also, a {\it signed\/} arithmetic
register whose range of values is known to lie within $[-2^n,+2^n)$
can be naturally represented in ``two's complement'' form by $n+1$
bits, and it is well known that arithmetic operations on such a two's
complement integer can be simulated by interpreting it as an integer
in the range $[0,2^{n+1})$ and performing arithmetic modulo $2^{n+1}$
(see, for example, \cite{Koren}).

\subsection{Addition and Conditional Addition}
\label{subsec:a}

In view of the above discussion, to simulate

\ni\ii {\bf if $B$ then $X \- X + k$},

\ni it suffices to perform

\ni\ii $X \- (X + B \cdot k) \bmod 2^n$

\ni (in other words to add $k$ to $X$ modulo $2^n$ iff $B = 1$).
In the case where $X$ is an $n$-bit signed register, it suffices to
substitute $n+1$ for $n$ above.

The program below performs this using $n$ auxiliary bits $C_0, C_1,
\ldots C_{n-1}$ (which are assumed to have initial value 0, and are
reset to 0 by the end of the computation).

\ni {\bf Program CONDITIONAL\_ADD\_$k$} \nl
\ii {\bf quantum registers:} \nl
\ii $X$ : $n$-bit signed arithmetic register \nl
\ii $B$ : bit register \nl
\ii $C_0,C_1,\ldots,C_{n-1}$ : bit registers
(initialized and finalized to 0) \nl
\nl
\ii {\bf for $i=1$ to $n-1$ do} \nl
\ii \ii $C_i \carrow C_i \xor \maj(k_{i-1},X_{i-1},C_{i-1})$ \nl
\ii {\bf for $i=n-1$ down to $1$ do} \nl
\ii \ii $X_i \- X_i \xor (k_i \wedge B)$ \nl
\ii \ii $X_i \- X_i \xor (C_i \wedge B)$ \nl
\ii \ii $C_i \crarrow C_i \xor \maj(k_{i-1},X_{i-1},C_{i-1})$ \nl
\ii $X_0 \- X_0 \xor k_0$

\ni where
$$\maj(l,S,T) = \cases{S \wedge T & if $l=0$ \cr
                       S \vee T & if $l=1$.\cr}$$

\ni The number of basic operations performed by the above program is
bounded above by $4n + O(1)$.  In particular, if the {\bf for} loops
of this program are unravelled then the program corresponds to a
gate-array consisting of $2n + 1$ bits and $4n + O(1)$ gates.  (A more
space-efficient $(n + O(\sqrt{n}))$-bit program is described in
Appendix \ref{ap:pebble}.

The unconditional addition statement

\ni\ii $X \- X + k$

\ni can be easily simulated by replacing $(k_i \wedge B)$ and
$(C_i \wedge B)$ in the above program with $k_i$ and $C_i$ (respectively).

{\bf CONDITIONAL\_ADD} introduces two modified assignment symbols
``$\carrow$'' and ``$\crarrow$''.  For the present purposes these can
be thought of as identical to the ordinary ``$\-$'' assignment;
however, they signal a freedom in how the quantum phase may be handled
in these assignments, as discussed in Appendix \ref{ap:phase}.

One final note about {\bf CONDITIONAL\_ADD}: it involves only the
addition of a quantum register with an ordinary, classical number.  It
is possible to write a similar program which adds two quantum
registers, as has been illustrated in \cite{Vedral}; however, this
more complex routine is never needed for the implementation of the
Schumacher function.  Actually, it is generally possible to implement
a full quantum adder as a sequence of calls to {\bf CONDITIONAL\_ADD}.

\subsection{Equality and Inequality Testing}
\label{subsec:b}

In order to simulate the remaining types of statements, it suffices to
simulate {\it equality test\/} statements of the form

\ni\ii $B \- B \xor (X = k)$

\ni (which negate $B$ iff $X = k$), and
{\it inequality test\/} statements of the form

\ni\ii $B \- B \xor (X > k)$

\ni (which negate $B$ iff $X > k$).

With implementations of the above tests, the statement

\ni\ii {\bf if $Y > l$ then $X \- X + k$}

\ni is then easily simulated by the sequence

\ni
\ii $B \carrow B \xor (Y > l)$ \nl
\ii {\bf if $B$ then $X \- X + k$} \nl
\ii $B \crarrow B \xor (Y > l)$

\ni where $B$ is a bit register distinct from the bits of $X$ and $Y$,
and whose initial value is 0 (note that $B$ must be reset to 0 after
the addition is performed).  Also, the compound conditional

\ni\ii {\bf if $Y = l$ and $B$ then $X \- X + k$}

\ni is simulated by the sequence

\ni
\ii $C \carrow C \xor (Y = l)$ \nl
\ii $D \carrow D \xor (C \wedge B)$ \nl
\ii {\bf if $D$ then $X \- X + k$} \nl
\ii $D \crarrow D \xor (C \wedge B)$ \nl
\ii $C \crarrow C \xor (Y = l)$

\ni where $C$ and $D$ are bit registers distinct from the bits of $X$,
$Y$, and $B$, and whose initial (and final) values are 0.  Again, the
meaning and usefulness of the phase-modified assignments is discussed
in Appendix \ref{ap:phase}.

The following program simulates an equality test.  It uses $n$
auxiliary bit registers $C_0,C_1,\ldots,C_{n-1}$.  The auxiliary
registers are initialized to 0, and have final value 0.

\ni {\bf Program TEST\_EQUALITY\_TO\_$k$} \nl
\ii {\bf quantum registers:} \nl
\ii $X$ : $n$-bit signed arithmetic register \nl
\ii $B$ : bit register \nl
\ii $C_0,C_1,\ldots,C_{n-1}$ : bit registers
(initialized and finalized to 0) \nl
\nl
\ii $C_{n-1} \carrow C_{n-1} \xor (X_{n-1} = k_{n-1})$ \nl
\ii {\bf for $i=n-2$ down to $0$ do} \nl
\ii \ii $C_i \carrow C_i \xor (C_{i+1} \wedge (X_i = k_i))$ \nl
\ii $B \- B \xor C_0$ \nl
\ii {\bf for $i=0$ to $n-2$ do} \nl
\ii \ii $C_i \crarrow C_i \xor (C_{i+1} \wedge (X_i = k_i))$ \nl
\ii $C_{n-1} \crarrow C_{n-1} \xor (X_{n-1} = k_{n-1})$

\ni where
$$(S = l) = \cases{\overline{S} & if $l=0$ \cr
                    S           & if $l=1$.\cr}$$

\ni The number of basic operations performed by the above program is
bounded above by $2n + O(1)$.  (The above program is very similar to
the so-called $\wedge_n$-gate construction in \cite{G9}).

Finally, the following program simulates an inequality test.  It uses
$n$ auxiliary bit registers $C_0,C_1,\ldots,C_{n-1}$.  The auxiliary
registers are initialized to 0, and have final value 0.

\ni {\bf Program TEST\_GREATER\_THAN\_$k$} \nl
\ii {\bf quantum registers:} \nl
\ii $X$ : $n$-bit signed arithmetic register \nl
\ii $B$ : bit register \nl
\ii $C_0,C_1,\ldots,C_{n-1}$ : bit registers
(initialized and finalized to 0) \nl
\nl
\ii $C_{n-1} \carrow C_{n-1} \xor (X_{n-1} = k_{n-1})$ \nl
\ii $B \- B \xor (X_{n-1} < k_{n-1})$ \nl
\ii {\bf for $i=n-2$ down to $0$ do} \nl
\ii \ii $C_i \carrow C_i \xor (C_{i+1} \wedge (X_i = k_i))$ \nl
\ii \ii $B \ciarrow B \xor C_{i+1} \wedge (X_i > k_i)$ \nl
\ii {\bf for $i=0$ down to $n-2$ do} \nl
\ii \ii $C_i \crarrow C_i \xor (C_{i+1} \wedge (X_i = k_i))$ \nl
\ii $C_{n-1} \crarrow C_{n-1} \xor (X_{n-1} = k_{n-1})$

\ni where $(S = l)$ is as in the previous subsection,
$$(S > l) = \cases{S & if $l=0$ \cr
                   0 & if $l=1$,\cr}$$
and
$$(S < l) = \cases{0            & if $l=0$ \cr
                   \overline{S} & if $l=1$.\cr}$$

\ni The number of basic operations performed by the above program is
bounded above by $3n + O(1)$.  Once again, we employ phase-modified
assignments $\carrow$, $\crarrow$, and $\ciarrow$ which are explained
in Appendix \ref{ap:phase}.

\section{Discussion and Conclusions}
\label{sec:conc}

We can finally put all the above results together to evaluate the
total cost, in time and space, to perform Schumacher coding.  It is
easy to see that the two {\bf if} statements inside the $i$ loop of
{\bf FINAL\_SCHUMACHER} are the most expensive part of the procedure.
The first {\bf if} statement requires one call to {\bf
CONDITIONAL\_ADD}.  Although $X$ is an $n$-bit register, the addition
only affects the $n-p-1$ low-order bits of $X$.  Thus, the addition
can be performed on $\mbox{\bf TRUNC}_{n-p-1}(X)$ rather than $X$,
which amounts to a total running time of
\begin{equation}
\sum_{p=0}^{n-1}\sum_{i=0}^{n-p}4(n-p-1)+O(1)=\mbox{$\frac{2}{3}$}n^3+O(n^2).
\end{equation}
The expensive part of the second {\bf if} statement is its two calls to
{\bf TEST\_GREATER\_\\-THAN}, performed on an $(n-p-1)$-bit quantum
register (because of the action of {\bf TRUNC}).  The time involved
for this is
\begin{equation}
\sum_{p=0}^{n-1}\sum_{i=0}^{n-p}2\cdot3(n-p-1)+O(1)=n^3+O(n^2).
\end{equation}
Thus, the total time required (i.e., number of bit-level primitive
steps) is $\frac{5}{3}n^3+O(n^2)$.  The total number qubits used is:
$n$, to hold the input/output string $X$; plus $\lceil \log n \rceil$,
to hold $S$; plus $n+O(1)$ to implement the conditional additions and
inequality tests (the same work registers that store carries and so
forth may be reused throughout the execution of the program).
Thus, the total number of qubits is $2n + \lceil \log n \rceil + O(1)$.

If the space-efficient routines {\bf CONDITIONAL\_ADD$^{\prime}$} and {\bf
TEST\_GREATER\_\\-THAN$^{\prime}$} introduced in Appendix \ref{ap:pebble}
are used instead, the execution time is increased to
$\frac{8}{3}n^3 +O(n^{2.5})$,
but the total number of qubits is reduced to
$n + 2\sqrt{n} + O(\log n)$.
If the relevant figure of merit for the tractability of the quantum
computation is the product of time and space, as it is in certain
physical models\cite{Unruh,Palma}, then the space-efficient procedures
we have introduced would be preferred.

A final note about these operation counts: they are all in terms of
the primitive operations listed at the beginning of Section
\ref{sec:bit}, which includes both two- and three-bit primitives.  It
is known\cite{G9,Detal} that all three-bit operations can be simulated
in quantum logic by a sequence of two-bit primitives.  Most of the
three-bit operations can be simulated using seven operations (3
quantum XORs and 4 one-bit gates); see Appendix \ref{ap:phase}.  So,
in terms of these primitive operations the total time to do the
Schumacher function would be roughly $7\cdot\frac{8}{3}n^3 < 19n^3$.
Computing the exact prefactor would require a considerable amount of
detailed calculation, and would have to take into account that fact
that many one-bit gates in the network could be merged together and
executed in one step (see \cite {G9}).  All of this work could easily
be done if an actual physical implementation of Schumacher compression
were ever undertaken.

To conclude, we believe that the pseudocode in which our results are
presented is the most concise and economical form in which to present
a quantum computation like Schumacher coding.  The bit level
primitives for addition and comparison which we have presented are
similar to ones which have been presented elsewhere\cite{Vedral}, but
have a few features which may make them superior in the development of
other quantum programs.  The Schumacher coding can be done in $O(n^3)$
steps, with $O(\sqrt{n})$ auxiliary workspace.  We cannot exclude the
possibility that a lower polynomial-order algorithm may be found, but
we are not presently aware of what form this would take.  The
techniques in \cite{Bennett} enable further shrinkage of the auxiliary
workspace, but with a larger penalty in the running time.  We think
that further useful shrinkage of the auxiliary workspace is unlikely;
in the present scheme, only a vanishingly small fraction of quantum
bits are used as workspace for large blocksize $n$.

\acknowledgements

We are grateful to C. H. Bennett, R. Jozsa, B. Schumacher, and J. Smolin
for useful discussions.  Thanks to the authors of Ref. \cite{Vedral} for
a preview of their work prior to publication.
R.C. is supported in part by NSERC of Canada.

\appendix

\section{Improvements in the Workspace Efficiency of the Bit-Level
Implementations}
\label{ap:pebble}

The bit-level implementations proposed in Sections \ref{subsec:a} and
\ref{subsec:b} require $n$ auxiliary bit registers.  By applying
techniques that were introduced in \cite{Bennett}, we derive the
following alternate programs that employ only $O(\sqrt{n})$ auxiliary
bit registers while maintaining the same asymptotic operation
complexity.  (The space-reduction techniques in \cite{Bennett}, can
also be used to reduce the auxiliary space further, but this incurs
an increase in the running time, as well as in the space-time
product.)

Assume that $n = m^2$.  The program {\bf
CONDITIONAL\_ADD$^{\prime}$\hspace*{-1mm}\_$k$} that follows employs
$2m-1$ auxiliary bit registers rather than the $n$ auxiliary bit
registers that {\bf CONDITIONAL\_ADD\_$k$} employs.  In {\bf
CONDITIONAL\_ADD\_$k$}, registers $C_0,\ldots,C_{n-1}$ are used to
store information about carry propagation.  In {\bf
CONDITIONAL\_ADD$^{\prime}$\hspace*{-1mm}\_$k$}, this is accomplished
by registers $C_1,\ldots,C_{m-1}$ and $D_1,\ldots,D_{m-1}$ instead.
The idea is to reset some of the registers to 0 at various checkpoints
during the course of the computation.  This is illustrated by the
diagram below, where the horizontal direction represents time, and the
placement of the lines indicate the time intervals during which the
registers are active, containing the various carry bits.  Registers
$D_1,\ldots,D_{m-1},C_1$ are first set to the first $m$ carry bits.
Then $D_1,\ldots,D_{m-1}$ are reset to 0.  Registers
$D_1,\ldots,D_{m-1},C_2$ can then be used to store the
$m+1^{\mbox{\scriptsize st}}$ to $2m^{\mbox{\scriptsize th}}$ carry
bits and then $D_1,\ldots,D_{m-1}$ are reset to 0 again --- since
$C_1$ stores the $m^{\mbox{\scriptsize th}}$ carry bit, this can be
accomplished without recomputing the first $m$ carry bits.  The
process is repeated with the remaining carry bits, and then applied in
reverse to reset the carry bits to 0, as illustrated here:

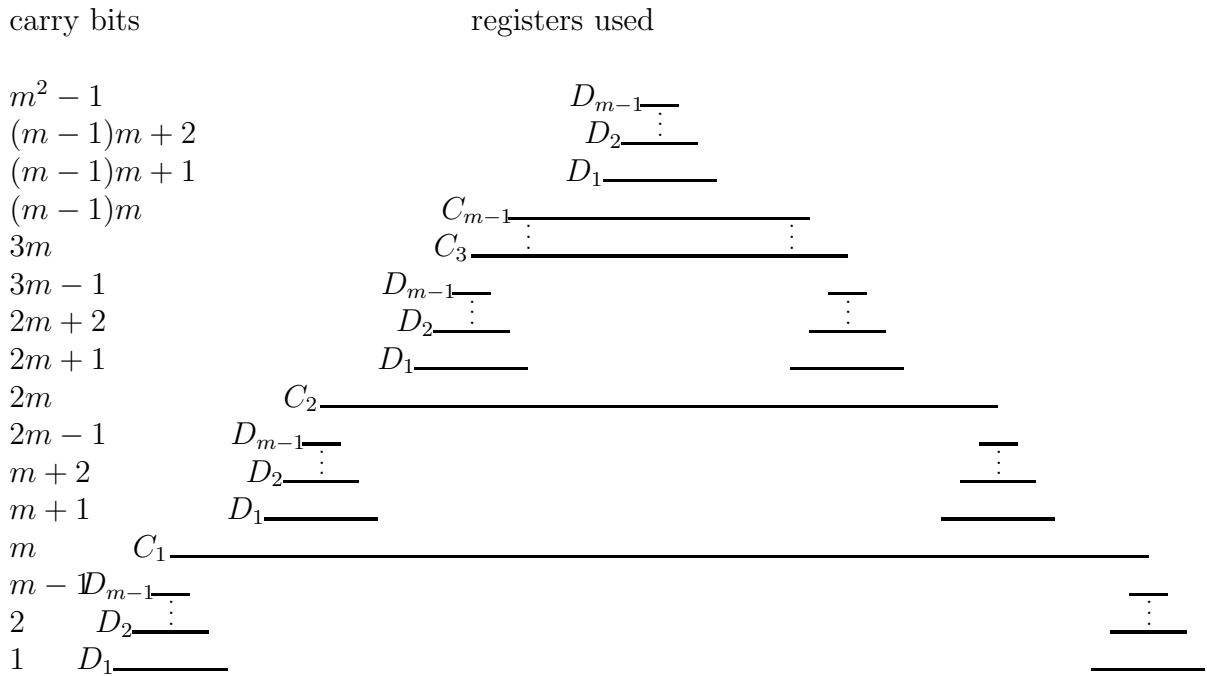
\setlength{\unitlength}{0.5cm}
\begin{picture}(2,19)(0,1)
\put(0,18){carry bits}
\put(0,1){$\scriptsize 1$}
\put(0,2){$\scriptsize 2$}
\put(0,3){$\scriptsize m-1$}
\put(0,4){$\scriptsize m$}
\put(0,5){$\scriptsize m+1$}
\put(0,6){$\scriptsize m+2$}
\put(0,7){$\scriptsize 2m-1$}
\put(0,8){$\scriptsize 2m$}
\put(0,9){$\scriptsize 2m+1$}
\put(0,10){$\scriptsize 2m+2$}
\put(0,11){$\scriptsize 3m-1$}
\put(0,12){$\scriptsize 3m$}
\put(0,13){$\scriptsize (m-1)m$}
\put(0,14){$\scriptsize (m-1)m+1$}
\put(0,15){$\scriptsize (m-1)m+2$}
\put(0,16){$\scriptsize m^2-1$}
\end{picture}
\begin{picture}(29,19)(0,1)

\thicklines

\put(10,18){registers used}
\put(0.5,1){\line(1,0){3}}
\put(1,2){\line(1,0){2}}
\put(1.5,3){\line(1,0){1}}

\put(2,4){\line(1,0){26}}

\put(26.5,1){\line(1,0){3}}
\put(27,2){\line(1,0){2}}
\put(27.5,3){\line(1,0){1}}

\put(-0.5,1.0){$\scriptsize D_1$}
\put(0.0,2.0){$\scriptsize D_2$}
\put(-0.4,3.0){$\scriptsize D_{m-1}$}
\put(1.0,4.0){$\scriptsize C_1$}

\put(4.5,5){\line(1,0){3}}
\put(5,6){\line(1,0){2}}
\put(5.5,7){\line(1,0){1}}

\put(6,8){\line(1,0){18}}

\put(22.5,5){\line(1,0){3}}
\put(23,6){\line(1,0){2}}
\put(23.5,7){\line(1,0){1}}

\put(3.5,5.0){$\scriptsize D_1$}
\put(4.0,6.0){$\scriptsize D_2$}
\put(3.6,7.0){$\scriptsize D_{m-1}$}
\put(5.0,8.0){$\scriptsize C_2$}

\put(8.5,9){\line(1,0){3}}
\put(9,10){\line(1,0){2}}
\put(9.5,11){\line(1,0){1}}

\put(10,12){\line(1,0){10}}

\put(18.5,9){\line(1,0){3}}
\put(19,10){\line(1,0){2}}
\put(19.5,11){\line(1,0){1}}

\put(7.5,9.0){$\scriptsize D_1$}
\put(8.0,10.0){$\scriptsize D_2$}
\put(7.6,11.0){$\scriptsize D_{m-1}$}
\put(9.0,12.0){$\scriptsize C_3$}

\put(11,13){\line(1,0){8}}

\put(13.5,14){\line(1,0){3}}
\put(14,15){\line(1,0){2}}
\put(14.5,16){\line(1,0){1}}

\put(9.2,13.0){$\scriptsize C_{m-1}$}

\put(12.5,14.0){$\scriptsize D_1$}
\put(13.0,15.0){$\scriptsize D_2$}
\put(12.6,16.0){$\scriptsize D_{m-1}$}


\put(1.94,2.2){$\scriptsize \vdots$}
\put(27.94,2.2){$\scriptsize \vdots$}

\put(5.94,6.2){$\scriptsize \vdots$}
\put(23.94,6.2){$\scriptsize \vdots$}

\put(9.94,10.2){$\scriptsize \vdots$}
\put(19.94,10.2){$\scriptsize \vdots$}

\put(14.94,15.2){$\scriptsize \vdots$}

\put(11.44,12.2){$\scriptsize \vdots$}
\put(18.44,12.2){$\scriptsize \vdots$}


\end{picture}\vspace*{4mm}

\ni The detailed program follows.  $D_0$ is used for convenience to
store the value of $C_i$ at the beginning of each iteration of the
for-loop with respect to $i$.

\newpage

\ni {\bf Program CONDITIONAL\_ADD$^{\prime}$\hspace*{-1mm}\_$k$} \nl
\ii {\bf quantum registers:} \nl
\ii $X$ : $n$-bit arithmetic register \nl
\ii $B$ : bit register \nl
\ii $C_1,C_2,\ldots,C_{m-1}$ : bit registers
(initialized and finalized to 0) \nl
\ii $D_0,D_1,\ldots,D_{m-1}$ : bit registers
(initialized and finalized to 0) \nl
\nl
\ii {\bf for $i=0$ to $m-2$ do} \nl
\ii \ii {\bf if $i > 0$ then $D_0 \- D_0 \xor C_i$} \nl
\ii \ii {\bf for $j=1$ to $m-1$ do} \nl
\ii \ii \ii $D_j \carrow D_j \xor \maj(k_{im+j-1},X_{im+j-1},D_{j-1})$ \nl
\ii \ii $C_{i+1} \carrow C_{i+1} \xor \maj(k_{im+m-1},X_{im+m-1},D_{m-1})$ \nl
\ii \ii {\bf for $j=m-1$ down to $1$ do} \nl
\ii \ii \ii $D_j \crarrow D_j \xor \maj(k_{im+j-1},X_{im+j-1},D_{j-1})$ \nl
\ii \ii {\bf if $i > 0$ then $D_0 \- D_0 \xor C_i$} \nl
\nl
\ii $D_0 \- D_0 \xor C_{m-1}$ \nl
\ii {\bf for $j=1$ to $m-1$ do} \nl
\ii \ii $D_j \carrow D_j \xor \maj(k_{(m-1)m+j-1},X_{(m-1)m+j-1},D_{j-1})$ \nl
\ii {\bf for $j=m-1$ down to $1$ do} \nl
\ii \ii $X_{(m-1)m+j} \- X_{(m-1)m+j} \xor (k_{(m-1)m+j} \wedge B)$ \nl
\ii \ii $X_{(m-1)m+j} \- X_{(m-1)m+j} \xor (D_j \wedge B)$ \nl
\ii \ii $D_j \crarrow D_j \xor \maj(k_{(m-1)m+j-1},X_{(m-1)m+j-1},D_{j-1})$ \nl
\ii $D_0 \- D_0 \xor C_{m-1}$ \nl
\nl
\ii {\bf for $i=m-2$ down to $0$ do} \nl
\ii \ii {\bf if $i > 0$ then $D_0 \- D_0 \xor C_i$} \nl
\ii \ii {\bf for $j=1$ to $m-1$ do} \nl
\ii \ii \ii $D_j \carrow D_j \xor \maj(k_{im+j-1},X_{im+j-1},D_{j-1})$ \nl
\ii \ii $X_{im+m} \- X_{im+m} \xor (k_{im+m} \wedge B)$ \nl
\ii \ii $X_{im+m} \- X_{im+m} \xor (C_{i+1} \wedge B)$ \nl
\ii \ii $C_{i+1} \crarrow C_{i+1} \xor \maj(k_{im+m-1},X_{im+m-1},D_{m-1})$ \nl
\ii \ii {\bf for $j=m-1$ down to $1$ do} \nl
\ii \ii \ii $X_{im+j} \- X_{im+j} \xor (k_{im+j} \wedge B)$ \nl
\ii \ii \ii $X_{im+j} \- X_{im+j} \xor (D_j \wedge B)$ \nl
\ii \ii \ii $D_j \crarrow D_j \xor \maj(k_{im+j-1},X_{im+j-1},D_{j-1})$ \nl
\ii \ii {\bf if $i > 0$ then $D_0 \- D_0 \xor C_i$} \nl
\nl
\ii $X_0 \- X_0 \xor k_0$

\ni The above program uses $n + O(\sqrt{n})$ registers in total and runs
in $6n + O(\sqrt{n})$ steps (compared to $2n + O(\log n)$ registers in
total and $4n + O(1)$ steps for {\bf CONDITIONAL\_ADD\_$k$}).

There also exist more space-efficient versions of
{\bf TEST\_EQUALITY\_TO\_$k$} and {\bf TEST\_GREATER\_THAN\_$k$}.
For the former case, the program is as follows (where again $n = m^2$).\ee

\ni {\bf Program TEST\_EQUALITY\_TO$^{\prime}$\hspace*{-1mm}\_$k$} \nl
\ii {\bf quantum registers:} \nl
\ii $X$ : $n$-bit arithmetic register \nl
\ii $B$ : bit register \nl
\ii $C_0,C_1,\ldots,C_{m-1}$ : bit registers
(initialized and finalized to 0) \nl
\ii $D_1,D_2,\ldots,D_{m}$ : bit registers
(initialized and finalized to 0) \nl
\nl
\ii {\bf for $i=m-1$ down to $0$ do} \nl
\ii \ii {\bf if $i=m-1$ then $D_m \- D_m \xor 1$
                        else $D_m \- D_m \xor C_{i+1}$} \nl
\ii \ii {\bf for $j=m-1$ down to $1$ do} \nl
\ii \ii \ii $D_j \carrow D_j \xor (D_{j+1} \wedge (X_{im+j} = k_{im+j}))$ \nl
\ii \ii $C_{i} \carrow C_{i} \xor (D_{1} \wedge (X_{im} = k_{im}))$ \nl
\ii \ii {\bf for $j=1$ to $m-1$ do} \nl
\ii \ii \ii $D_j \crarrow D_j \xor (D_{j+1} \wedge (X_{im+j} = k_{im+j}))$ \nl
\ii \ii {\bf if $i=m-1$ then $D_m \- D_m \xor 1$
                        else $D_m \- D_m \xor C_{i+1}$} \nl
\nl
\ii $B \- B \xor C_0$ \nl
\nl
\ii {\bf for $i=0$ to $m-1$ do} \nl
\ii \ii {\bf if $i=m-1$ then $D_m \- D_m \xor 1$
                        else $D_m \- D_m \xor C_{i+1}$} \nl
\ii \ii {\bf for $j=m-1$ down to $1$ do} \nl
\ii \ii \ii $D_j \carrow D_j \xor (D_{j+1} \wedge (X_{im+j} = k_{im+j}))$ \nl
\ii \ii $C_{i} \crarrow C_{i} \xor (D_{1} \wedge (X_{im} = k_{im}))$ \nl
\ii \ii {\bf for $j=1$ to $m-1$ do} \nl
\ii \ii \ii $D_j \crarrow D_j \xor (D_{j+1} \wedge (X_{im+j} = k_{im+j}))$ \nl
\ii \ii {\bf if $i=m-1$ then $D_m \- D_m \xor 1$
                        else $D_m \- D_m \xor C_{i+1}$} \ee

\ni The above program uses $n + O(\sqrt{n})$ registers in total and runs
in $4n + O(\sqrt{n})$ steps (compared to $2n + O(1)$ registers in
total and $2n + O(1)$ steps for {\bf TEST\_EQUALITY\_TO\_$k$}).

The program {\bf TEST\_GREATER\_THAN$^{\prime}$\hspace*{-1mm}\_$k$}
is similar to the above attaining $5n + O(\sqrt{n})$ time and
$n + O(\sqrt{n})$ space (vs.\ $3n + O(1)$ time and  $2n + O(1)$ space) .

\section{Phase Freedom in Implementation of Reversible Routines}
\label{ap:phase}

Here we will explain ways in which the quantum phase can be treated in
the essentially classical reversible routines which we have been
discussing throughout this paper.  In the language of quantum logic
gates, the bit-level logic statements used in the programs here are
represented by unitary matrices applied to the quantum wavefunction of
all the registers.  These unitary matrices have a special restriction
which make them ``classical", which is that the matrix elements are
only zero or one; this means that every definite computational state
$|x\rangle$ is taken to another definite computational state
$|f(x)\rangle$, and not to a superposition of states.  To give an
example, the Toffoli gate, the three-bit implementation of the AND
gate in reversible logic, involves the following unitary matrix:
\begin{equation}
\left (\begin{array}{rrrrrrrr}
\;\;1&\;\;0&\;\;0&\;\;0&\;\;0&\;\;0&\;\;0&\;\;0\\
\;\;0&\;\;1&\;\;0&\;\;0&\;\;0&\;\;0&\;\;0&\;\;0\\
\;\;0&\;\;0&\;\;1&\;\;0&\;\;0&\;\;0&\;\;0&\;\;0\\
\;\;0&\;\;0&\;\;0&\;\;1&\;\;0&\;\;0&\;\;0&\;\;0\\
\;\;0&\;\;0&\;\;0&\;\;0&\;\;1&\;\;0&\;\;0&\;\;0\\
\;\;0&\;\;0&\;\;0&\;\;0&\;\;0&\;\;1&\;\;0&\;\;0\\
\;\;0&\;\;0&\;\;0&\;\;0&\;\;0&\;\;0&\;\;0&\;\;1\\
\;\;0&\;\;0&\;\;0&\;\;0&\;\;0&\;\;0&\;\;1&\;\;0
\end{array}\right ).\label{noph}
\end{equation}

Here we consider the question of whether elementary operations with
modified phases (i.e., in which the matrix elements are unimodular
complex numbers $e^{i\theta}$, rather than being 1) could be used in
the implementation of the Schumacher function.  We are motivated to
investigate this because we found in our previous study\cite{G9} that
the implementation of a modified Toffoli gate with a single non-zero
phase
\begin{equation}
\left (\begin{array}{rrrrrrrr}
\;\;1&\;\;0&\;\;0&\;\;0&\;\;0&\;\;0&\;\;0&\;\;0\\
\;\;0&\;\;1&\;\;0&\;\;0&\;\;0&\;\;0&\;\;0&\;\;0\\
\;\;0&\;\;0&\;\;1&\;\;0&\;\;0&\;\;0&\;\;0&\;\;0\\
\;\;0&\;\;0&\;\;0&\;\;1&\;\;0&\;\;0&\;\;0&\;\;0\\
\;\;0&\;\;0&\;\;0&\;\;0&-1&\;\;0&\;\;0&\;\;0\\
\;\;0&\;\;0&\;\;0&\;\;0&\;\;0&\;\;1&\;\;0&\;\;0\\
\;\;0&\;\;0&\;\;0&\;\;0&\;\;0&\;\;0&\;\;0&\;\;1\\
\;\;0&\;\;0&\;\;0&\;\;0&\;\;0&\;\;0&\;\;1&\;\;0
\end{array}\right ),\label{cheap}
\end{equation}
requires fewer resources in the following sense: we showed that the
zero-phase Toffoli gate (Eq. (\ref{noph})) can be implemented with 8
two-bit XOR gates and 8 one-bit gates, while the Toffoli gate with
modified phases (Eq. (\ref{cheap})) requires only 3 XOR's and 4
one-bit gates.  We will establish here that the less-costly gate can
in fact be used for most of the Toffoli gates, and related three-bit
operations, that are used in the implementation of the Schumacher
compression function.

Note that it is necessary that the complete Schumacher calculation be
carried out with all the quantum phases equal to zero, in order that
the superposition states discussed in Section \ref{sec:one} maintain
the correct phase relation to one another.  Thus the question becomes:
how can the effect of the non-zero phase in Eq. (\ref{cheap}), if it
is introduced in one Toffoli gate, be undone at some later step of the
calculation?  The answer (which we will establish shortly) is the
obvious one: many of the reversible routines which we have introduced
(although not the high-level Schumacher program itself) have a
palindromic character, so that a Toffoli gate on three bits is exactly
``undone" at a later stage of the computation, roughly as far from the
end of the subroutine as the original gate is from the beginning.  It
turns out that the effect of the $-1$ phase factor can be precisely
undone at the second occurrence of the gate, too.

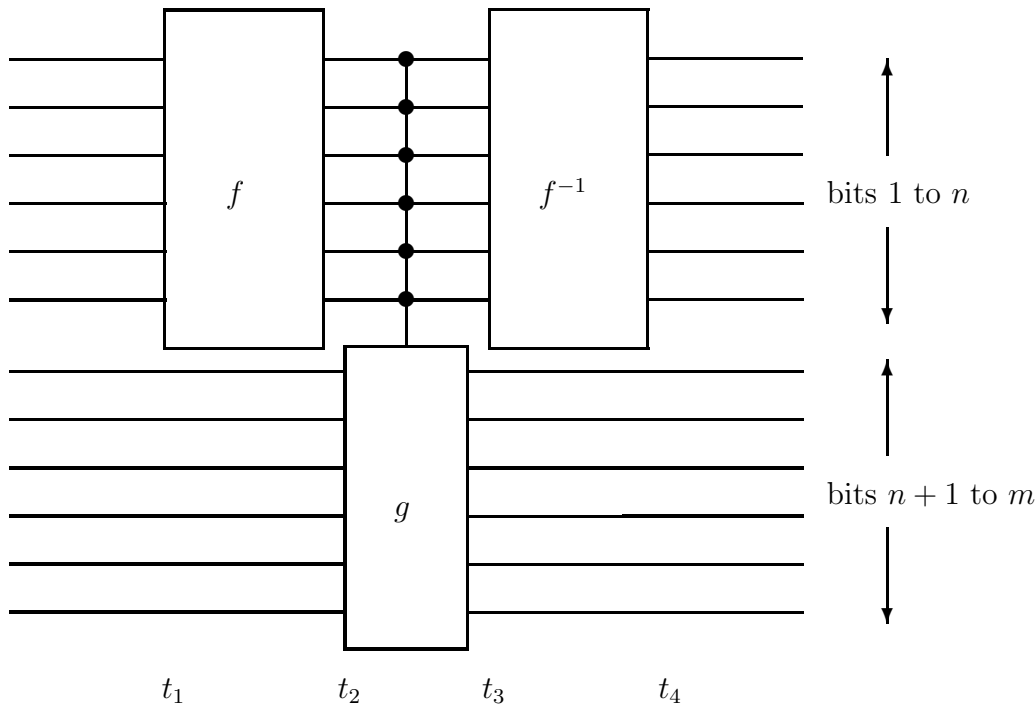
\setlength{\unitlength}{0.00083300in}%
\begingroup\makeatletter\ifx\SetFigFont\undefined%
\gdef\SetFigFont#1#2#3#4#5{%
  \reset@font\fontsize{#1}{#2pt}%
  \fontfamily{#3}\fontseries{#4}\fontshape{#5}%
  \selectfont}%
\fi\endgroup%
\begin{picture}(5648,4303)(1490,-5100)
\thicklines
\put(6452,-2310){\line(-1, 0){975}}
\put(6452,-2010){\line(-1, 0){975}}
\put(6451,-1709){\line(-1, 0){975}}
\put(6451,-1409){\line(-1, 0){975}}
\put(6451,-1109){\line(-1, 0){975}}
\put(1502,-1112){\line( 1, 0){975}}
\put(1502,-1412){\line( 1, 0){975}}
\put(1502,-1712){\line( 1, 0){975}}
\put(1503,-2013){\line( 1, 0){975}}
\put(1503,-2313){\line( 1, 0){975}}
\put(1503,-2613){\line( 1, 0){975}}
\put(1503,-3963){\line( 1, 0){975}}
\put(2478,-3963){\line( 1, 0){1123}}
\put(3601,-3961){\line(-1, 0){1125}}
\put(1503,-4563){\line( 1, 0){2098}}
\put(1503,-4263){\line( 1, 0){2098}}
\put(1502,-3662){\line( 1, 0){2099}}
\put(1502,-3362){\line( 1, 0){2099}}
\put(1502,-3062){\line( 1, 0){2099}}
\put(4354,-3964){\line( 1, 0){975}}
\put(5329,-3964){\line( 1, 0){1123}}
\put(6452,-3962){\line(-1, 0){1125}}
\put(4354,-4564){\line( 1, 0){2098}}
\put(4354,-4264){\line( 1, 0){2098}}
\put(4353,-3663){\line( 1, 0){2099}}
\put(4353,-3363){\line( 1, 0){2099}}
\put(4353,-3063){\line( 1, 0){2099}}
\put(6452,-2610){\line(-1, 0){975}}
\put(2851,-2010){$f$}
\put(3901,-3961){$g$}
\put(4801,-2010){$f^{-1}$}
\put(3976,-1111){\circle*{100}}
\put(3976,-1411){\circle*{100}}
\put(3976,-1711){\circle*{100}}
\put(3976,-2011){\circle*{100}}
\put(3976,-2311){\circle*{100}}
\put(3976,-2611){\circle*{100}}
\put(4501,-2909){\framebox(975,2100){}}
\put(3601,-4786){\framebox(750,1875){}}
\put(3453,-2613){\line( 1, 0){1048}}
\put(3453,-2313){\line( 1, 0){1048}}
\put(3453,-2013){\line( 1, 0){1048}}
\put(3452,-1712){\line( 1, 0){1049}}
\put(3452,-1412){\line( 1, 0){1049}}
\put(3452,-1112){\line( 1, 0){1049}}
\put(3976,-1111){\line( 0,-1){1800}}
\put(6976,-3586){\vector( 0, 1){600}}
\put(6976,-4036){\vector( 0,-1){600}}
\put(6976,-1711){\vector( 0, 1){600}}
\put(6976,-2161){\vector( 0,-1){600}}
\put(2476,-2911){\framebox(975,2100){}}
\put(6601,-2011){bits $1$ to $n$}
\put(6601,-3886){bits $n+1$ to $m$}
\put(2450,-5100){$t_1$}
\put(3550,-5100){$t_2$}
\put(4450,-5100){$t_3$}
\put(5550,-5100){$t_4$}
\end{picture}

We will now establish the desired basic result using the setup of the
figure, that the boolean function $f$ can be implemented with any
arbitrary phase factors, so long as they also appear in $f^{-1}$, no
matter what the intervening boolean function $g$ is, so long as $g$
does not modify the values of the bits on which $f$ and $f^{-1}$ act.
By applying this result repeatedly to the subroutines which we have
introduced, starting at the innermost level, we deduce all the
three-bit primitives which can be implemented with non-zero phase.
Assignments in which these non-zero phases are permitted have been
identified by the special assignment symbol
\begin{equation}
\carrow.\nonumber
\end{equation}
These statements are always paired with others, denoted by
\begin{equation}
\crarrow,\nonumber
\end{equation}
in which the reverse phases are implemented.  (For the phases in
Eq. (\ref{cheap}), the operation is self-inverse.)  In one case, the
pairing is between statements in different, palindromically-arranged
calls of the same subroutine; for these we have used a distinct symbol
\begin{equation}
\ciarrow.\nonumber
\end{equation}
After establishing the basic result, we will review a few of the
details of this implementation in the Schumacher function.

Let us first write down what the set of operations in the figure is
supposed to do.  Beginning with the basis state
\begin{equation}
|x\rangle=|\overbrace{x_1x_2...x_n...x_m}^{\mbox{$m$ bits}}\rangle
\end{equation}
at time $t_1$, it becomes at $t_2$, after the operation of $f$,
\begin{equation}
|x'\rangle=|\overbrace{f(x_1x_2...x_n)}^{\mbox{$n$ bits}}x_{n+1}...x_m\rangle
\end{equation}
Then at time $t_3$ the state is
\begin{equation}
\exp(i\theta_g(x'))
|\overbrace{f(x_1x_2...x_n)}^{\mbox{$n$ bits}}\!\!\!\!\!\!\!
\underbrace{g(x')}_{\mbox{$m-n$ bits}}\!\!\!\!\!\!\!\rangle.
\end{equation}
$g(x')$ depends on the state of the entire $m$-bit register $x'$, but
only modifies the last $m-n$ bits, as indicated.  Note that we allow
for the possibility that $g$ itself is a modified boolean function
with non-zero phases.  This is necessary because we will apply this
result in a nested fashion in the Schumacher subroutines.  Finally at
time $t_4$ the state is
\begin{equation}
\exp(i\theta_g(x))|x_1x_2...x_ng(x')\rangle.\label{sfinal}
\end{equation}
That is, the first $n$ bits are restored to their original state, and
bits $n+1$ through $m$ remain in the state $g(x)$.

Now, the question is, will the state Eq. (\ref{sfinal}) still result
if the function $f$ is modified to introduce non-zero phases
$\theta_f(x_1...x_n)$?  If we establish that this is true for all
boolean inputs $|x\rangle$, this will suffice to prove that these
networks have the same action on any arbitrary quantum states (this
follows directly from the linear superposition principle of quantum
mechanics).  We follow the time evolution as before with the modified
$f$.  At time $t_2$ the state is
\begin{equation}
\exp(i\theta_f(x_1...x_n))|f(x_1x_2...x_n)x_{n+1}...x_m\rangle
\end{equation}
Then at time $t_3$:
\begin{equation}
\exp(i(\theta_g(x')+\theta_f(x_1...x_n)))|f(x_1x_2...x_n)g(x')\rangle
\end{equation}
and finally at $t_4$:
\begin{equation}
\exp(i(\theta_f(x_1...x_n)+\theta_g(x')+\theta_{f^{-1}}(f(x_1...x_n))))
|x_1x_2...x_ng(x')\rangle\label{scheap}
\end{equation}
The final term in the phase factor can be simplified.  Recall that the
unitary transformation corresponding to $f^{-1}$ is the transpose of
the complex conjugate of the unitary transformation corresponding to
$f$.  (This follows directly from the definition of unitarity.)
Therefore, to get $\theta_{f^{-1}}$ from $\theta_f$, we flip the sign
(this is the complex conjugation), and we make the argument of the
$\theta$ function the {\em output} values of the bits rather than the
{\em input} values (this is the transpose).  Here we use the fact that
$g$ does not modify the first $n$ bits --- their output values are the
same as the original inputs $x_1x_2...x_n$.  Rendering this in
mathematical language:
\begin{equation}
\theta_{f^{-1}}(f(x_1...x_n))=-\theta_f(x_1...x_n)
\end{equation}
Thus, the two $\theta_f$ terms in the phase in Eq. (\ref{scheap})
cancel out, and Eq. (\ref{scheap}) becomes identical to
Eq. ({\ref{sfinal}), which is the desired result.$\Box$

Finally, we briefly review the application of this result to the
programs introduced in the text.  The first appearance is in {\bf
CONDITIONAL\_ADD\_$k$}, where the role of $f$ is played in the
innermost part of the program by the assignment statement

\ni \nl \ii \ii $C_{n-1} \carrow C_{n-1} \xor
\maj(k_{n-2},X_{n-2},C_{n-2})$ \nl

\ni This is a three-bit operation of the Toffoli type (or a trivial
modification of it) involving the bits $C_{n-1}$, $C_{n-2}$, and
$X_{n-2}$.  $f^{-1}$ occurs a short distance down,

\ni \nl \ii \ii $C_{n-1} \crarrow C_{n-1} \xor \maj(k_{n-2},X_{n-2},C_{n-2})$
\nl

\ni The role of $g$ is played by the two statements

\ni \nl \ii \ii $X_{n-1} \- X_{n-1} \xor (k_{n-1} \wedge B)$ \nl
\ii \ii $X_{n-1} \- X_{n-1} \xor (C_{n-1} \wedge B)$ \nl

\ni Obviously, only $X_{n-1}$ is modified by $g$, so the condition
that $g$ modify only bits not touched by $f$ is satisfied; so, we are
allowed to introduce a phase-modified $f$ as indicated by the
$\carrow$ and $\crarrow$ assignments.  Moving away from the innermost
part of the program, we see that all the above is nested inside a
larger $g$ in which $C_{n-1}$, $X_{n-2}$, and $X_{n-1}$ are modified,
surrounded by a $f-f^{-1}$ pair involving the bits $C_{n-2}$,
$X_{n-3}$, and $C_{n-3}$; working outward in succession this way, we
conclude that all the $C_i$ assignments may be replaced with
phase-modified $\carrow$ and $\crarrow$ assignments.

In Section \ref{subsec:b} we exhibit a pair of statements

\ni
\ii $B \carrow B \xor (Y > l)$ \nl
\ii $B \crarrow B \xor (Y > l)$

\ni playing the role of $f$ and $f^{-1}$.  These are not primitive
three-bit operations as in the earlier examples, but they are
themselves implemented with bit-level programs ({\bf
TEST\_GREATER\_THAN\_$k$}).  For this $f^{-1}$, the $\crarrow$
assignment requires that the bit-level routine be run in the
time-reversed order.  This can be done since classically, this boolean
function is its own inverse.  In the time-inverted {\bf
TEST\_GREATER\_THAN\_$k$}, the $\carrow$'s and $\crarrow$'s should be
interchanged.  The $B$ assignment involving the symbol $\ciarrow$ in
this routine is special, in that it is paired with the same statement
in the time-reversed call to {\bf TEST\_GREATER\_THAN\_$k$}.  This
special symbol is a reminder is that this statement should be
implemented with the phases corresponding to the $\carrow$ assignment
in the first call to the program, and with those corresponding to the
$\crarrow$ assignment in the second call.

We have not indicated phase-modifying assignments for any of the
two-bit gate level operations in these programs.  We take as given
that these two-bit gates could be implemented with zero phases.  But
if this were not the case, then many of these paired assignments may
be phase-modified in exactly the way we have shown for the three-bit
primitives.

\end{document}